# Novel Node-Based BAT for Finding All Minimal Cuts

Wei-Chang Yeh
Integration and Collaboration Laboratory
Department of Industrial Engineering and Engineering Management
National Tsing Hua University
yeh@ieee.org

*Abstract* — The binary-state network, a basic network, and its components are either working or failed. It is fundamental to all types of current networks, such as utility networks (gas, water, electricity, and 4G/5G), the Internet of Things (IoT), social networks, and supply chains. Network reliability is an important index in the planning, design, evaluation, and control of network systems, and the minimal path (MC) is the basis of an MC-based algorithm for calculating reliability. A new BAT called recursive node-based BAT is proposed and implemented in a recursive manner such that the *j*th vector in the *i*th iteration is equal to its parent, which is the *j*th vector, except that its *i*th coordinate value is one. Based on time complexity analysis and experiments on 20 benchmark binary-state networks, the proposed recursive node-based BAT is more efficient than the best-known node-based algorithm in finding MCs when combined with powerful rules to discard these infeasible vectors and all their offspring.

*Keywords*: Binary-state Network; Network Reliability; Minimal Cut (MC); Binary-Addition-Tree Algorithm (BAT); recursive method

## 1. INTRODUCTION

Modern society is full of various networks for transmitting, transferring, and/or transporting flow [1], gas [2], petrol [3], power [4, 5], vehicles [6], products [7-8], signals [9], data [10], multimedia [11], social relationship [12], etc., to have a better, more convenient, and efficient living life. Network reliability is defined as the probability of success of the current functions, states, and performance of the network [13, 14, 15]. Therefore, network reliability has been a popular index for measuring network functions, states, and performance for many decades [13, 14, 15].

Regardless of how complicated and delicate the networks are, their fundamental is the binary-state network, in which each component has a binary state: working or failed [16, 17, 18, 19, 20].

Relaxing the number of states of the binary-state networks results in multi-state flow networks (MFNs) that satisfy the flow conservation law [21, 22, 23, 24, 25], deterioration/augmentation-effect MFNs (DMFNs/AMFNs) [26, 27], that are based on pseudo-flow conservation, and multi-state information networks (MINs) that do not follow the flow conservation law [28, 29].

Allowing different kinds of flows, such as text, audio, and image data in cloud computing, MFNs, DMFNs/AMFNs, and MINs are generalized to multi-commodity MFNs [30] and MINs [31], respectively. Furthermore, multi-distribution networks allow each component to have more than one state distribution, and they extend multi-commodity MFNs to multi-distribution multi-commodity MFNs [32]. Thus, the improvement in the binary-state network is advantageous for all varieties of network models described above [16–32].

It is NP-Hard and #P-hard to calculate binary-state network reliability problems [13] with or without constraints, such as cost, volume, weight, $k$-out-of-$n$ [22, 23, 25], limited transmission speed [6], limited memory capacity, signal quality [9], and failure rate [33]. Real-life networks face various limitations and challenges [13, 14, 15]. Consequently, conditional network reliability problems are more important than these problems without accounting for constraints [34].

For conditional binary-state network reliability problems, all algorithms focus on either minimal cuts (MCs) [18, 19, 20, 23] or minimal paths (MPs) [1, 2, 3, 4]. Both the MP and MC are arc subsets [5, 7, 8, 9, 18, 19]. An MP is a simple path from the source node to the sink node, whereas an MC is a cut in which its removal separates the source and sink nodes [18, 19]. None of the arcs in MCs or MPs are redundant; otherwise, the MC/MP is a cut/path rather than a real MC/MP [13, 14].

The numbers of MPs and MCs are $2^m$ and $2^n$ in the worst case, where $n$ is the number of nodes and $m = O(n^2)$ is the number of arcs [13, 14]. Hence, among the MP algorithms that found all MPs and MC algorithms that found all MCs, the MC algorithms were more efficient than the MP algorithms [18, 19]. Thus, this study focused only on finding MCs.

It is also NP-Hard and #P-hard to find all MCs or MPs [13]. Both MC and MP algorithms are indirect algorithms because they cannot calculate the network reliability directly and need to implement another method, such as the inclusion-exclusion method [35] or sum-of-disjoint product [36], to use MPs and MCs to achieve the final goal. Unfortunately, calculating the binary-state network reliability using the inclusion-exclusion method or sum-of-disjoint product in terms of these found MCs or MPs is NP-Hard and #P-hard [13]. Therefore, determining all MCs is an important topic in MC algorithms.

Various MC algorithms have been proposed to determine all MCs before calculating the reliability of various networks [18, 19, 20, 23]. Among these algorithms, the node-based MC algorithm [19] is the most efficient because it finds all MCs based on nodes rather than arcs, and the time complexities of the former and latter are $O(2^n)$ and $O(2^m)$ [13], respectively.

However, the node-based MC algorithm is based on the depth-first-search (DFS) [18, 19, 20, 23], which is less efficient than the BAT proposed by Yeh [37, 38]. BAT is a straightforward implicit enumeration method that can efficiently generate all possible required vectors or solutions. Recently, the BAT was improved by Yeh using a recursive method in [39], such that the $j$th vector in the $i$th iteration is equal to its parent, which is the $j$th vector, except that its $i$th coordinate value is one. The time complexity required to obtain a new $n$-tuple $X$ is reduced from $O(n)$ in traditional BAT to $O(1)$ in recursive BAT.

Recursive BAT can also reduce the number of infeasible vectors by discarding each infeasible vector as soon as it is found [39]. The most useful aspect is that recursive BAT can further avoid these infeasible vectors generated from infeasible vectors in the recursive procedure.

To the best of our knowledge, no recursive MC algorithm for identifying all MCs has been reported to date. Hence, the goal of this study is to propose a recursive node-based algorithm based on recursive BAT together with some new concepts to avoid having infeasible vectors to reduce the number of infeasible vectors and to reduce the time required to verify whether a vector is feasible for efficiently calculating the binary-state network reliability.

The remainder of this study is organized as follows. All acronyms, notations, nomenclature, and assumptions required in the proposed recursive node-based BAT for finding all MCs are defined in Section 2. A short review of MCs, BAT, PLSA, and IET is presented in Section 3. The best-known node-based MC algorithm is implemented based on the BAT in Section 4. The details of the major innovations of the proposed algorithm in reducing the number of infeasible vectors and decreasing the runtime to verify the feasibility of vectors are presented in Section 5. The proposed algorithm is formally introduced in Section 6, together with a discussion of the pseudocode, time complexity, and demonstrations using a step-by-step example. The performance of the proposed algorithm was further validated by comparing it with the best-known node-based MC algorithm, conducted on 20 binary-state benchmark networks. Finally, Section 7 concludes the study and discusses possible future work.

## 2. ACRONYMS, NOTATIONS, NOMENCLATURE, AND ASSUMPTIONS

All required acronyms, notations, nomenclature, and assumptions are given in this section.

### 2.1 Acronyms

MC: Minimal cut

MP: Minimal path

IET: Inclusion–exclusion technology

BAT: Binary-addition tree algorithm [37, 38]

### 2.2 Notations

$|\bullet|$: Number of elements in set $\bullet$

$n$: Number of nodes

$n^*$: Number of nontarget nodes without including nodes 1 and $n$, i.e., $n^* = n - 2$

$m$: Number of arcs

$c$: Number of MCs

$V$: Set of nodes $V = \{1, 2, \ldots, n\}$

$E$: Set of arcs $E = \{a_1, a_2, \ldots, a_m\}$

$a_k$: $k^{\text{th}}$ arc in $E$

$e_{i,j}$: undirected arc from node $i$ to node $j$

$G(V, E)$ : A graph with $V$, $E$, source node 1, and sink node $n$; for example, Figure 1 is a graph with $V = \{1, 2, \ldots, 7\}$, $E = \{a_1, a_2, \ldots, a_{12}\}$, source node 1, and sink node 4.

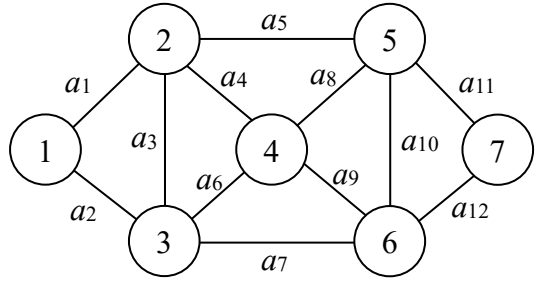

**Figure 1.** Example network.

$\mathbf{D_b}$ : State distribution lists the probability for each working arc; e.g., $\mathbf{D_b} = \{\Pr(a_1) = \Pr(a_3) = \Pr(a_5) = \Pr(a_7) = \Pr(a_9) = \Pr(a_{11}) = 0.96, \Pr(a_2) = \Pr(a_4) = \Pr(a_6) = \Pr(a_8) = \Pr(a_{10}) = \Pr(a_{12}) = 0.91\}$ in Figure 1.

$G(V, E, \mathbf{D_b})$ : Binary-state network with $G(V, E)$ and $\mathbf{D_b}$. For example, Figure 1 with $\mathbf{D_b} = \{\Pr(a_1) = \Pr(a_3) = \Pr(a_5) = \Pr(a_7) = \Pr(a_9) = \Pr(a_{11}) = 0.96, \Pr(a_2) = \Pr(a_4) = \Pr(a_6) = \Pr(a_8) = \Pr(a_{10}) = \Pr(a_{12}) = 0.91\}$ is a binary-state network.

$X$ : $n^*$-tuple node-based vector such that its $i$th coordinate is represented whether node ($i+1$) is in the subnetwork with node 1 after the related cut. For example, $X_{24} = (1, 1, 1, 0, 1)$ denotes nodes 2, 3, 4, and 6 are in the subnetwork with node 1 after removing the MC $C(X_{24}) = \{a_5, a_8, a_{10}, a_{12}\}$ related to $X$ as shown in Figure 2, where $C(X_{24})$ is defined below.

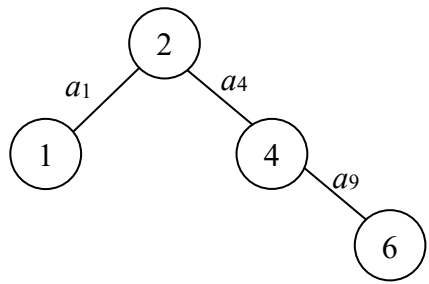

**Figure 2.** $G(X_{22})$, where $X_{22} = (1, 0, 1, 0, 1)$ in Figure 1.

$X(v)$ : $v$th coordinate value represented whether node ($v+1$) is in the subnetwork with node 1 after the related cut $C(X)$. For example, $X_{22}(1) = X_{22}(3) = X_{22}(5) = 1$ in $X_{22} = (1, 0, 1, 0, 1)$.

$S(X)$: Node subset $S(X) = \{ v \in V \mid X(v) = 1\} \cup \{1\}$. For example, $S(X_{22}) = \{1, 2, 4, 6\}$ (see Figure 2) if $X_{22} = (1, 0, 1, 0, 1)$ in Figure 1.

$T(X)$: Node subset $T(X) = \{ v \in V \mid X(v) = 0\} \cup \{n\} = V - S(X)$. For example, $T(X_{22}) = \{3, 5, 7\}$ if $X_{22} = (1, 0, 1, 0, 1)$ in Figure 1.

$C(X)$: Arc subset $C(X) = \emptyset$ and $\{ e_{i,j} \in V \mid$ for all $i \in S(X)$ and $j \in T(X)\}$ if $X$ is infeasible and feasible, respectively. For example, $C(X_{22}) = \emptyset$ and $C(X_{24}) = \{a_5, a_8, a_{10}, a_{12}\}$ because $X_{22} = (1, 0, 1, 0, 1)$ is infeasible and $X_{24} = (1, 1, 1, 0, 1)$ is feasible in Figure 1.

$G(X)$: Subgraph $G(X) = G(X, E)$ related to vector node-based $X$ such that node $v \in G(X)$ for all $X(v) = 1$ and node $a \in G(X)$ if two endpoints of $a$ are in $G(X)$. For example, $G(X_{22})$ shown in Figure 2 is a subgraph of Figure 1 where $X_{22} = (1, 0, 1, 0, 1)$.

$V(v)$: Node subset $V(v) = \{ v \in V \mid e_{v,j} \in E$ for all $j \in V\}$. For example, $V(2) = \{1, 3, 4, 5\}$ in Figure 1.

$L(i)$: Node subset in the $i$th layer based on the PLSA of which its details are given in Section 3.3. For example, $L(2) = \{2, 3\}$ in Figure 1.

$\lambda(v)$: Index denoted which layer of node $v$. For example, $\lambda(2) = \lambda(3) = 2$ if $L(2) = \{2, 3\}$.

$U(X)$: Node subset $U(X) = \emptyset$ and $\{ v \in T(X) \mid e_{i,v} \in C(X)$ for all $i \in S(X) \}$ if $X$ is infeasible and feasible, respectively. For example, $U(X_{22}) = \emptyset$ and $U(X_{24}) = \{5, 7\}$ because $X_{22} = (1, 0, 1, 0, 1)$ is infeasible and $X_{24} = (1, 1, 1, 0, 1)$ is feasible in Figure 1.

$\Pr(\bullet)$: Probability to have $\bullet$ successfully, i.e., $\Pr(X) = \prod_{X(a)=1} \Pr(a)$

$\mathrm{supp}(\bullet_k)$: $\mathrm{supp}(\bullet_k) = \{X \mid X \geq X_k\}$, where $X_k(a_i) = \begin{cases} 1 & \text{if } a_i \in \bullet_k \\ 0 & \text{otherwise} \end{cases}$. Note that $\mathrm{supp}(\bullet_j \cap \bullet_k) = \mathrm{supp}(\bullet_j) \cup \mathrm{supp}(\bullet_k)$.

## 2.3 Nomenclature

Reliability: The connected probability between nodes 1 and $n$.

MC: A special cut between nodes 1 and $n$ without any redundant arc of which its removal without affecting to MCs. For example, $\{a_4, a_5, a_6, a_7\}$ is an MC, $\{a_4, a_5, a_6\}$ is neither a cut nor an MC, and $\{a_2, a_4, a_5, a_6, a_7\}$ is a cut with a redundant arc $a_2$, i.e., not an MC, in Figure 1.

MP: A simple path from nodes 1 to $n$ without any redundant arc. For example, $\{a_2, a_7, a_{12}\}$ is an MP, $\{a_2, a_7\}$ is neither a path nor an MP, and $\{a_1, a_2, a_7, a_{12}\}$ is an MP with a redundant arc $a_1$ in Figure 1.

Isolated node: A node $v$ is an isolated in $X$ if it cannot be reached from node 1 and to node $n$ in $G(S(X))$ and $G(T(X))$, respectively. For example, node 3 is an isolated node in $X_{22} = (1, 0, 1, 0, 1)$ because $3 \in T(X_{22})$ and 3 cannot reach node 7 in $G(T(X_{22}))$ from Figure 1.

Edge node: Each node is called an edge node in $X$ if it is in $U(X)$. For example, both nodes 5 and 7 are edge nodes in $X_{24} = (1, 1, 1, 0, 1)$ from Figure 1.

Feasible Vector: A node-based vector $X$ is feasible, and $C(X)$ is an MC if no isolated nodes in $X$. For example, in Figure 1, $X_{22} = (1, 0, 1, 0, 1)$ and $X_{24} = (1, 1, 1, 0, 1)$ are infeasible and feasible because node 3 is an isolated node in $X_{22}$ and $X_{24}$ has no isolated nodes, respectively.

## 2.4 Assumptions

1. Each node, say $v$, is perfectly reliable, and at least one simple path is from nodes 1 to $n$ via $v$.
2. Each arc is either functioning or has failed, and its occurrence probability is statistically independent, with a predetermined distribution from observation, empirical data, or testing.
3. No loops and/or parallel arcs exist.

## 3. OVERVIEW OF MC, BAT, PLSA, AND IET

Before introducing the proposed algorithm for finding MCs and calculating the binary-state network reliability problem in terms of MCs, the basic concepts, including MC, BAT, PLSA, and IET, behind the proposed algorithm are briefly discussed.

### 3.1 MC and MC-Based Algorithms

An MC is a special cut whose removal separates the network into two subnetwork nodes, such that nodes 1 and $n$ are not in the same subnetwork [19]. The difference between cut and MC is that each arc in an MC is not redundant, and the removal of any arc in an MC will result in the MC not being cut. For example, $C = \{C_1, C_2, C_3, C_4\}$ is a complete MC set in Figure 3, where $C_1 = \{a_1, a_2\}$, $C_2 = \{a_1, a_3, a_5\}$, $C_3 = \{a_2, a_3, a_4\}$, and $C_4 = \{a_4, a_5\}$. The arc subset $C^* = \{a_1, a_2, a_3\}$ is only a cut, not an MC, because $a_3$ is redundant and can be removed from $C^*$.

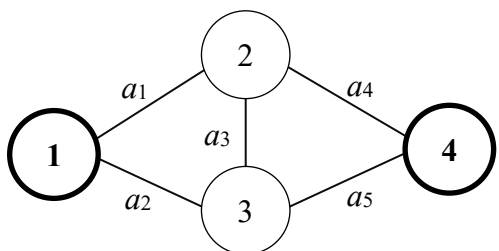

**Figure 3.** Example network.

Various methods have been proposed for searching MCs owing to their importance in solving binary-state network reliability problems. All methods are based on implicit enumeration algorithms heuristic algorithms [18], such as DFS [19, 20], and universal generating function methodologies [40].

Among these methods, the node-based MC algorithm proposed in [19] is the best-known. The node-based MC algorithm provides a new concept for finding all MCs from arc to node subsets. Because the number of arcs is the square of that of nodes in general, the node-MC algorithm is more efficient than these arc-based MC algorithms in finding all MCs. The BAT-based node-MC algorithm is described in Section 3.4.

### 3.2 BAT

The arc-based BAT, first developed by Yeh [37], is based on the following two rules to generate all possible binary-state vectors:

**Rule 1.** The coordinate $a_i$ with the first zero value is changed to one, and $a_j = 0$ for $j < i$.

**Rule 2.** If $a_i = 1$ for all $i$, halt and all vectors are found.

The above two rules of BAT are described in pseudocode as follows [41]:

**Algorithm 1: Arc-based BAT**

**Input:**    $k$.

**Output:**   All $k$-tuple arc-based binary-state vectors.

**STEP 0.**   Let $X$ be a $k$-tuple vector 0 and $i = 1$ be the current index of coordinate in $X$.

**STEP 1.**   If $X(a_i) = 0$, let $X(a_i) = 1$, $i = 1$, and go to STEP 1.

**STEP 2.**   Halt if $i = k$.

**STEP 3.**   Let $X(a_i) = 0$, $i = i + 1$, and return STEP 1.

The BAT is you-only-look-once (YOLO), such that only vector $X$ is updated repeatedly in the 4-line BAT pseudo-code without saving it. From the above, BAT has four lines and one $X$. As a result, BAT is simple to code, efficient to run, memory-friendly, and easy to customize. Thus, BAT has been implemented in many different applications, such as resilience appraisal [42], spread probability of wildfires [43], propagation probability of computer viruses [44], and various network reliability problems [45-49]

For example, in Figure 3, all vectors $X$ obtained from the above pseudocode are listed in Table 1. Note that the index $j$ of $X_j$ makes it easy for the reader to recognize which iteration has the corresponding $X$, and it is not needed in the BAT.

**Table 1.** All the $X_i$ values obtained in the proposed BAT.

| $j$ | $X_j$ | $j$ | $X_j$ |
|---|---|---|---|
| 1 | (0, 0, 0, 0, 0) | 17 | (1, 0, 0, 0, 0) |
| 2 | (0, 0, 0, 0, 1) | 18 | (1, 0, 0, 0, 1) |
| 3 | (0, 0, 0, 1, 0) | 19 | (1, 0, 0, 1, 0) |
| 4 | (0, 0, 0, 1, 1) | 20 | (1, 0, 0, 1, 1) |
| 5 | (0, 0, 1, 0, 0) | 21 | (1, 0, 1, 0, 0) |
| 6 | (0, 0, 1, 0, 1) | 22 | (1, 0, 1, 0, 1) |
| 7 | (0, 0, 1, 1, 0) | 23 | (1, 0, 1, 1, 0) |
| 8 | (0, 0, 1, 1, 1) | 24 | (1, 0, 1, 1, 1) |
| 9 | (0, 1, 0, 0, 0) | 25 | (1, 1, 0, 0, 0) |
| 10 | (0, 1, 0, 0, 1) | 26 | (1, 1, 0, 0, 1) |
| 11 | (0, 1, 0, 1, 0) | 27 | (1, 1, 0, 1, 0) |
| 12 | (0, 1, 0, 1, 1) | 28 | (1, 1, 0, 1, 1) |
| 13 | (0, 1, 1, 0, 0) | 29 | (1, 1, 1, 0, 0) |
| 14 | (0, 1, 1, 0, 1) | 30 | (1, 1, 1, 0, 1) |
| 15 | (0, 1, 1, 1, 0) | 31 | (1, 1, 1, 1, 0) |
| 16 | (0, 1, 1, 1, 1) | 32 | (1, 1, 1, 1, 1) |

### 3.3 Layers and PLSA

Network reliability is defined by the success-connected probability of the networks under some redefined conditions. Hence, in network reliability problems, it is always necessary to verify the connectivity of the obtained state vectors, regardless of the MC- or MP-based algorithms. To verify whether a vector is connected efficiently, the PLSA proposed in [37] and its variants [38, 47, 48, 49] are always combined with BAT to solve various network reliability problems, such as one-to-one [37, 38], all-pairs [45], many-to-many [47], all-level network reliability problems [48], Monte Carlo Simulation Problems [49], and Redundancy Allocation Problems [50].

The PLSA originates from the layered-search algorithm (LSA) [51], which builds many disjoint node subsets, $L_1, L_2, \ldots, L_l$, called layers in LSA consecutively, such that each node in $L_i$ has at least one arc adjacent to one node in $L_{i+1}$ and not $L_{i+j}$ for $j = 1, 2, \ldots, (l-i)$. Let node $s$ be the only node in $L_1$: If node $t$ is in a layer different from $L_1$, it concludes that nodes $s$ and $t$ are connected. The PLSA pseudocode is as follows.

**Algorithm 2: PLSA**

**Input:** $G(V, E)$ and vector $X$.

**Output:** Whether nodes 1 and $n$ is connected in $G(X)$.

**STEP P0.** Let $L_1 = \{1\}$ and $i = 2$.

**STEP P1.** Let $L_i = \{ v \notin (L_1 \cup L_2 \cup \ldots \cup L_{i-1}) \mid$ for all $e_{\alpha,v} \in E$ and $\alpha \in L_{i-1}\}$.

**STEP P2.** If $L_i = \emptyset$, $X$ is disconnected and halt.

**STEP P3.** If $n \in L_i$, $X$ is connected and halt.

**STEP P4.** Let $i = i + 1$ and go to STEP P1.

Table 2. Process from the proposed PLSA for Figure 2.

| $i$ | $L_i$ | $L_{i+1}$ |
|---|---|---|
| 1 | {1} | {2} |
| 2 | {2} | {4} |
| 3 | {4} | {6} |
| 4 | {6} | |

The time complexity of the PLSA is only $O(n)$ because each node is at most included in one layer. Hence, PLSA is simple and efficient. For example, in Figure 1, the procedure to determine whether $S(X_{22})$ is connected is shown in Figure 2, where $X_{22} = (1, 0, 1, 0, 1)$.

**3.4 IET**

Because of its straightforward and significant role, IET is a popular algorithm in calculating network reliability in terms of MCs after having all MCs. IET calculates the binary-state network reliability based on the following equation [39]:

$$Pr(\bigcup_{i=1}^{p} supp(P_i)) = \sum_{k=1}^{p}(-1)^{k+1} Pr(\bigcap_{P_i \in I_k} supp(P_i)), \qquad (1)$$

where $I_k$ is the set of $k$ different MPs, $supp(P_k) = \{X \mid X \geq X_k\}$, $X_k$ is the $k$-th 1-MP, i.e., $X_k(a_i) = \begin{cases} 1 & \text{if } a_i \in P_k \\ 0 & \text{otherwise} \end{cases}$, $supp(P_k) \cap supp(P_l) = \{X \mid X \geq Max(X_k, X_l) = \begin{cases} 1 & \text{if } a_i \in P_k \cup P_l \\ 0 & \text{otherwise} \end{cases}\} = supp(P_k \cup P_l)$.

Each term on the right side of Eq. (1) is an intersection of MC subsets and is called the IET term. Probability based on Eq. (1) in terms of MCs is calculated as follows:

$$R^{\#}(G) = Pr(C_1) + Pr(C_2) + Pr(C_3) + Pr(C_4)$$
$$- [Pr(C_1 \cap C_2) + Pr(C_1 \cap C_3) + Pr(C_1 \cap C_4) + Pr(C_2 \cap C_3) + Pr(C_2 \cap C_4) + Pr(C_3 \cap C_4)]$$
$$+ [Pr(C_1 \cap C_2 \cap C_3) + Pr(C_1 \cap C_2 \cap C_4) + Pr(C_1 \cap C_3 \cap C_4) + Pr(C_2 \cap C_3 \cap C_4)]$$
$$- Pr(C_1 \cap C_2 \cap C_3 \cap C_4). \qquad (2)$$

$$= Pr(\{a_1, a_2\}) + Pr(\{a_1, a_3, a_5\}) + Pr(\{a_2, a_3, a_4\}) + Pr(\{a_4, a_5\})$$
$$- [Pr(\{a_1, a_2, a_3, a_5\}) + Pr(\{a_1, a_2, a_3, a_4\}) + Pr((\{a_1, a_2, a_4, a_5\}) + Pr(\{a_1, a_2, a_3, a_4, a_5\})$$
$$+ Pr(\{a_1, a_3, a_4, a_5\}) + Pr(\{a_2, a_3, a_4, a_5\})]$$
$$+ [Pr(\{a_1, a_2, a_3, a_4, a_5\}) + Pr(\{a_1, a_2, a_3, a_4, a_5\}) + Pr(\{a_1, a_2, a_3, a_4, a_5\}) + Pr(\{a_1, a_2, a_3, a_4, a_5\})]$$
$$- Pr(\{a_1, a_2, a_3, a_4, a_5\}). \qquad (3)$$

## 4 BAT-BASED NODE-BASED MC

The best-known node-based MC algorithm is modified for implementation in BAT to combine MC, MC algorithm, PLSA, and BAT in this section.

### 4.1 Pseudocode

The node-based MC algorithm was originally implemented based on the DFS [19]. From the experiments, BAT was found to be more efficient than DFS. A BAT-based node-based MC algorithm is first proposed, and its pseudocode is as follows:

**Algorithm 3: Proposed BAT-based Node-based MC Algorithm**

**Input:** $G(V, E)$, source node 1 and sink node $n$.

**Output:** All MCs.

**STEP M0.** Let $X = \mathbf{0}$, $u = 1$, $n^* = (n-2)$, and $\Omega = \{ \{a_{1,j} \mid a_{1,j} \in E\}\}$.

**STEP M1.** If $X(u) = 0$, let $X(u) = 1$, $u = 1$, and go to STEP M4.

**STEP M2.** If $u = n^*$, halt.

**STEP M3.** Let $X(u) = 0$, $u = u + 1$, and go to STEP M1.

**STEP M4.** If both $G(S(X))$ and $G(T(X))$ are connected subgraphs without isolated nodes, $C(X)$ is an MC and let $\Omega = \Omega \cup \{C(X)\}$. Go to STEP M1.

Basically, the above pseudocode implements the node-based MC algorithm in the original BAT. However, there are still two major differences.

1. $X$ is changed from an arc-based $m$-tuple vector to a node-based $n^*$-tuple.
2. STEP M4 is added to find the MC after having a new $X$.

The time complexity for the BAT to find all $n$-tuple binary-state vectors is $O(n2^n)$, and the time complexity to implement the PLSA in STEP 4 is $O(n)$ for each obtained vector. Hence, the proposed BAT-based node-based MC algorithm has a time complexity $O(n^2 2^n)$ in finding all MCs in binary-state networks.

## 4.2 Step-by-Step Example

The binary-state network shown in Figure 1 is utilized to explain the procedure of the proposed BAT-based node-based MC algorithm for finding all MCs step-by-step, as follows.

**STEP M0.** Let $X = \mathbf{0}$, $u = 1$, $n^* = (n-2) = 5$, and $\Omega = \{\{a_1, a_2\}\}$.

**STEP M1.** Because $X(u) = 0$, let $X(u) = 1$, $u = 1$, and go to STEP M4.

**STEP M4.** Because $G(S(X))$ and $G(T(X))$ both are connected, $C(X) = \{a_2, a_3, a_4, a_5\}$ is a MC, let $\Omega = \Omega \cup \{U(X)\} = \{\{a_1, a_2\}, \{a_2, a_3, a_4, a_5\}\}$, and go to STEP M1, where $S(X) = \bigcup_{\substack{v=1 \\ X(v)=1}}^{n} \{v\} \cup \{1\} = \{1, 2\}$ and $T(X) = \{3, 4, 5, 6, 7\}$.

$$\vdots$$

**STEP M1.** Because $X(3) = 0$ in $X = (1, 1, 0, 0, 0)$, let $X(3) = 1$, $u = 1$, and go to STEP M4, i.e., $X$ is updated to $(0, 0, 1, 0, 0)$.

**STEP M4.** Because $G(S(X))$ is disconnected, $C(X)$ is not an MC, and go to STEP M1, where $S(X) = \bigcup_{\substack{v=1 \\ X(v)=1}}^{n} \{v\} \cup \{1\} = \{1, 4\}$.

$$\vdots$$

**STEP M1.** Because $X(1) = 0$ in $X = (0, 0, 1, 0, 1)$, let $X(1) = 1$, $u = 1$, and go to STEP M4, i.e., $X$ is updated to $(1, 0, 1, 0, 1)$.

**STEP M4.** Because $G(S(X))$ is connected but $G(T(X))$ is disconnected, $C(X)$ is not a MC, and go to STEP M1, where $S(X) = \bigcup_{\substack{v=1 \\ X(v)=1}}^{n} \{v\} \cup \{1\} = \{1, 2, 4, 6\}$ and $T(X) = \{3, 5, 7\}$.

$$\vdots$$

**STEP M1.** Because $X(5) = 1$ in $X = (0, 0, 0, 0, 1)$, go to STEP M2.

**STEP M2.** Because $u = n^* = 5$, halt.

The whole procedure and results are summarized in Table 3, where vector $X$ is feasible if values of $U(X)$, $T(X)$, and $C(X)$ are listed.

**Table 3.** All $X_i$ and MCs obtained from the proposed node-based BAT.

| $i$ | $X_i$ | $S(X_i)$ | $U(X_i)$ | $T(X_i)$ | $C(X)$ |
|---|---|---|---|---|---|
| 1 | (0, 0, 0, 0, 0) | {1} | {2, 3} | {2, 3, 4, 5, 6, 7} | {$a_1$, $a_2$} |
| 2 | (1, 0, 0, 0, 0) | {1, 2} | {3, 4, 5} | {3, 4, 5, 6, 7} | {$a_2$, $a_3$, $a_4$, $a_5$} |
| 3 | (0, 1, 0, 0, 0) | {1, 3} | {2, 4, 6} | {2, 4, 5, 6, 7} | {$a_1$, $a_3$, $a_6$, $a_7$} |
| 4 | (1, 1, 0, 0, 0) | {1, 2, 3} | {4, 5, 6} | {4, 5, 6, 7} | {$a_4$, $a_5$, $a_6$, $a_7$} |
| 5 | (0, 0, 1, 0, 0) | {1, 4} | | | |
| 6 | (1, 0, 1, 0, 0) | {1, 2, 4} | {3, 5, 6} | {3, 5, 6, 7} | {$a_2$, $a_3$, $a_5$, $a_6$, $a_8$, $a_9$} |
| 7 | (0, 1, 1, 0, 0) | {1, 3, 4} | {2, 5, 6} | {2, 5, 6, 7} | {$a_1$, $a_3$, $a_4$, $a_7$, $a_8$, $a_9$} |
| 8 | (1, 1, 1, 0, 0) | {1, 2, 3, 4} | {5, 6} | {5, 6, 7} | {$a_5$, $a_7$, $a_8$, $a_9$} |
| 9 | (0, 0, 0, 1, 0) | {1, 5} | | | |
| 10 | (1, 0, 0, 1, 0) | {1, 2, 5} | {3, 4, 6, 7} | {3, 4, 6, 7} | {$a_2$, $a_3$, $a_4$, $a_8$, $a_{10}$, $a_{11}$} |
| 11 | (0, 1, 0, 1, 0) | {1, 3, 5} | | | |
| 12 | (1, 1, 0, 1, 0) | {1, 2, 3, 5} | {4, 6, 7} | {4, 6, 7} | {$a_4$, $a_6$, $a_7$, $a_8$, $a_{10}$, $a_{11}$} |
| 13 | (0, 0, 1, 1, 0) | {1, 4, 5} | | | |
| 14 | (1, 0, 1, 1, 0) | {1, 2, 4, 5} | {3, 6, 7} | {3, 6, 7} | {$a_2$, $a_3$, $a_6$, $a_9$, $a_{10}$, $a_{11}$} |
| 15 | (0, 1, 1, 1, 0) | {1, 3, 4, 5} | | | |
| 16 | (1, 1, 1, 1, 0) | {1, 2, 3, 4, 5} | {6, 7} | {6, 7} | {$a_7$, $a_9$, $a_{10}$, $a_{11}$} |
| 17 | (0, 0, 0, 0, 1) | {1, 6} | | | |
| 18 | (1, 0, 0, 0, 1) | {1, 2, 6} | | | |
| 19 | (0, 1, 0, 0, 1) | {1, 3, 6} | {2, 4, 5, 7} | {2, 4, 5, 7} | {$a_1$, $a_3$, $a_6$, $a_9$, $a_{10}$, $a_{12}$} |
| 20 | (1, 1, 0, 0, 1) | {1, 2, 3, 6} | {4, 5, 7} | {4, 5, 7} | {$a_4$, $a_6$, $a_7$, $a_8$, $a_{10}$, $a_{11}$} |
| 21 | (0, 0, 1, 0, 1) | {1, 4, 6} | | | |
| 22 | (1, 0, 1, 0, 1) | {1, 2, 4, 6} | | | |
| 23 | (0, 1, 1, 0, 1) | {1, 3, 4, 6} | {2, 5, 7} | {2, 5, 7} | {$a_4$, $a_5$, $a_6$, $a_9$, $a_{12}$} |
| 24 | (1, 1, 1, 0, 1) | {1, 2, 3, 4, 6} | {5, 7} | {5, 7} | {$a_5$, $a_8$, $a_{10}$, $a_{12}$} |
| 25 | (0, 0, 0, 1, 1) | {1, 5, 6} | | | |
| 26 | (1, 0, 0, 1, 1) | {1, 2, 5, 6} | | | |
| 27 | (0, 1, 0, 1, 1) | {1, 3, 5, 6} | | | |
| 28 | (1, 1, 0, 1, 1) | {1, 2, 3, 5, 6} | | | |
| 29 | (0, 0, 1, 1, 1) | {1, 4, 5, 6} | | | |
| 30 | (1, 0, 1, 1, 1) | {1, 2, 4, 5, 6} | | | |
| 31 | (0, 1, 1, 1, 1) | {1, 3, 4, 5, 6} | | | |
| 32 | (1, 1, 1, 1, 1) | {1, 2, 3, 4, 5, 6} | | {7} | {$a_{11}$, $a_{12}$} |

## 5. PROPOSED NOVEL CONCEPTS

The major difference between the proposed algorithm and that in Section 4 is the recursive method used to verify the connectivity of all obtained vectors. The details of this process are discussed in this section.

### 5.1 Renumber Nodes Based on PLSA

Let vector $X(v) = 0$ for all $v \geq u$. Vector $X^*$ is called the son of $X$ if $X^*(v) = \begin{cases} 1 & v = u \\ X(v) & \text{otherwise} \end{cases}$, and $X^{\#}$ is an offspring of $X$ if $X \ll X^{\#}$ and $X^{\#}(v) = X(v)$ for all $v < u$. For example, let $X_6 = (1, 0, 1, 0, 0)$. $X_{14} = (1, 0, 1, 1, 0)$ is the son of $X_6$ and $X_{14} = (1, 0, 1, 1, 0)$, $X_{22} = (1, 0, 1, 0, 1)$, and $X_{30} = (1, 0, 1, 1, 1)$ are offspring of $X_6$.

As shown in Table 3, $X_5 = (0, 0, 1, 0, 0)$ is infeasible, and all of its offspring, that is, $X_{13} = (0, 0, 1, 1, 0)$, $X_{21} = (0, 0, 1, 0, 1)$, and $X_{29} = (0, 0, 1, 0, 1)$, with the first three coordinates being equal to that in $X_5$, are also infeasible.

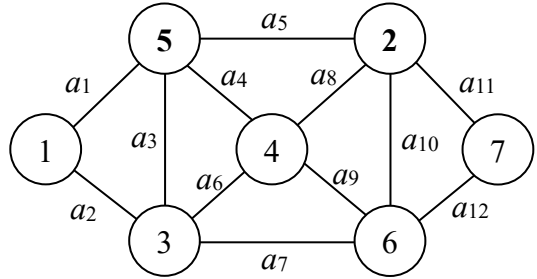

**Figure 4.** Example network with nodes 2 and 5 swapped in Figure 1.

However, this phenomenon is not always the case. For example, in the same network, nodes 2 and 5 were swapped, as shown in Figure 4. It is still true that $X_5 = (0, 0, 1, 0, 0)$ is infeasible, and $X_{13} = (0, 0, 1, 1, 0)$ is still one offspring of $X_5$. However, $X_{13} = (0, 0, 1, 1, 0)$ is feasible and completely different from that shown in Figure 1.

Let $X$ be infeasible and $X(i) = 0$ for $i = k+1, k+2, \ldots, n-2$. If $G(S(X))$ or $G(T(X))$ are still disconnected after letting $X(i) = 1$ for $i = k+1, k+2, \ldots, n-2$, we have that all offspring of $X$ are infeasible. Therefore, a new method based on PLSA is proposed to renumber nodes such that a vector is infeasible and that all offspring are infeasible. This method can skip the related infeasible offspring once we know that it is infeasible to reduce the number of infeasible vectors.

**Algorithm 4: PLSA-based Renumber Nodes**

**Input:** $G(V, E)$, source node 1 and sink node $n$.

**Output:** All MCs.

**STEP N0.** Implement PLSA to all nodes in $G(V, E)$.

**STEP N1.** Let $L(v)$ be the layer index obtained from the PLSA for all nodes $v \in V$.

**STEPN 2.** Renumber nodes such that nodes $i < j$ if nodes $i$ and $j$ are in $L_\iota$ and $L_\varphi$, respectively, with $\iota \leq \varphi$.

**STEP N3.** Renumber nodes in each layer such that nodes $i < j$ if nodes $L_s(i) > L_s(i)$.

For example, followed STEP 1, $L(i)$ for all $i$ and $\lambda(v)$ for each node $v$ in Figure 1 are shown in Table 4.

**Table 4.** All $X_i$ and MCs obtained from the proposed node-based BAT.

| $i$ | 1 | 2 | 3 | 4 | | | |
|---|---|---|---|---|---|---|---|
| $L(i)$ | {1} | {2, 3} | {4, 5, 6} | {7} | | | |
| $v$ | 1 | 2 | 3 | 4 | 5 | 6 | 7 |
| $\lambda(v)$ | 1 | 2 | 2 | 3 | 3 | 3 | 4 |

## 5.2 Recursive Concept

The recursive BAT was first proposed in [39] to implement the inclusion-exclusion method to calculate the network reliability after determining the connected vectors and disconnected vectors, that is, MPs and MCs, respectively. To efficiently verify the feasibility of the vector updated from a feasible vector, the proposed algorithm adapted the recursive BAT to easily have edge nodes.

In recursive BAT [39], there are $n^* = (n-2)$ iterations for the problem discussed here with $n^*$ variables. The $i$th iteration has $2^{(\iota-1)}$ vectors, where $\iota = n_{j-1}$ is the number of vectors in the $(i-1)$th iteration for $i = 2, 3, \ldots, n^*$ and the first iteration has 2 vectors. Each vector, say $X$, in the $i$th iteration with $X(j) = 0$ for $j = (i+1), (i+2), \ldots, n^*$. Hence, without loss of generality and for convenience, $X$ generated in the $i$th iteration is written in $i$-tuple because $X(j) = 0$ for all $j > i$.

In the recursive BAT, let $X_j$ be the $j$th vector generated, $X_{i,j}$ be the $j$th vector in the $i$th iterative for $j = 1, 2, \ldots, 2^{(\iota-1)}$, $\iota = n_{j-1}$, $i = 1, 2, \ldots, (m-2)$, $X_1 = X_{1,1} = (0)$, and $X_2 = X_{1,2} = (1)$. The following important equation is the core of recursive BAT to obtain all vectors:

$$X_{i,j}(k) = \begin{cases} X_j(k) & k = 1, 2, \ldots, (i-1) \\ 1 & \text{otherwise} \end{cases}. \tag{4}$$

Also,

$$S(X_{i,j}) = S(X_j) \cup \{(i+2)\}, \tag{5}$$

$$U(X_{i,j}) = [U(X_j) - \{(i+2)\}] \cup [V(i+2) - S(X_j)]. \tag{6}$$

Thus, the recursive BAT can have one vector in $O(1)$ [39], which is much faster than $O(n)$ in the traditional BAT [37].

For example, Tables 5-9 listed the results after implementing the recursive method for Algorithm 3 in Figure 1, where we have two vectors in both the first and second iterations, as shown in Tables 5 and 6, respectively. Note that $X_{i,j}$, and $X_j$ are listed in the same row and their differences are shown in bold.

**Table 5.** Results in the first iterative for vectors 1 and 2.

| $j$ | $X_j = X_{1,j}$ | $S(X_j)$ | $U(X_j)$ | $C(X_j)$ |
|---|---|---|---|---|
| 1 | (0) | {1} | {2, 3} | {$a_1, a_2$} |
| 2 | (1) | {1, 2} | {3, 4, 5} | {$a_2, a_3, a_4, a_5$} |

**Table 6.** Results in the 2nd iterative for vectors 3 and 4 based on vectors 1 and 2, respectively.

| $j$ | $X_j=X_{1,j}$ | $S(X_j)$ | $U(X_j)$ | $i$ | $j$ | $X_i=X_{2,j}$ | $S(X_i)$ | $U(X_i)$ | $C(X_i)$ |
|---|---|---|---|---|---|---|---|---|---|
| 1 | (0) | {1} | {2, 3} | 3 | 1 | (0,**1**) | {1, **3**} | {2, 4, 6} | {$a_1, a_3, a_6, a_7$} |
| 2 | (1) | {1, 2} | {3, 4, 5} | 4 | 2 | (1,**1**) | {1, 2, **3**} | {4, 5, 6} | {$a_4, a_5, a_6, a_7$} |

**Table 7.** Results in the 3rd iterative for vectors 5-8 based on that of 1-4, respectively.

| $i$ | $X_i$ | $S(X_i)$ | $U(X_i)$ | $i$ | $X_i$ | $S(X_i)$ | $U(X_i)$ | $C(X_i)$ |
|---|---|---|---|---|---|---|---|---|
| 1 | (0) | {1} | {2, 3} | 5 | (0,0,**1**) | {1, **4**}$^D$ | | |
| 2 | (1) | {1, 2} | {3, 4, 5} | 6 | (1,0,**1**) | {1, 2, **4**} | {4, 5, 6} | {$a_2, a_3, a_5, a_6, a_8, a_9$} |
| 3 | (0,1) | {1, 3} | {2, 4, 6} | 7 | (0,1,**1**) | {1, 3, **4**} | {2, 5, 6} | {$a_1, a_3, a_4, a_7, a_8, a_9$} |
| 4 | (1,1) | {1, 2, 3} | {4, 5, 6} | 8 | (1,1,**1**) | {1, 2, 3, **4**} | {5, 6} | {$a_5, a_7, a_8, a_9$} |

**Table 8.** Results in the 4th iterative for vectors 9-16 based on that of 1-8, respectively.

| $i$ | $X_i$ | $S(X_i)$ | $U(X_i)$ | $i$ | $X_i$ | $S(X_i)$ | $U(X_i)$ | $C(X_i)$ |
|---|---|---|---|---|---|---|---|---|
| 1 | (0) | {1} | {2, 3} | 9 | (0,0,0,**1**) | {1, **5**}$^D$ | | |
| 2 | (1) | {1, 2} | {3, 4, 5} | 10 | (1,0,0,**1**) | {1, 2, **5**} | {3, 4, 6, 7} | {$a_2, a_3, a_4, a_8, a_{10}, a_{11}$} |
| 3 | (0,1) | {1, 3} | {2, 4, 6} | 11 | (0,1,0,**1**) | {1, 3, **5**}$^D$ | | |
| 4 | (1,1) | {1, 2, 3} | {4, 5, 6} | 12 | (1,1,0,**1**) | {1, 2, 3, **5**} | {4, 6, 7} | {$a_4, a_6, a_7, a_8, a_{10}, a_{11}$} |
| 5 | (0,0,1) | {1, 4}$^D$ | | 13 | (0,0,1,**1**) | {1, 4, **5**}$^D$ | | |
| 6 | (1,0,1) | {1, 2, 4} | {4, 5, 6} | 14 | (1,0,1,**1**) | {1, 2, 4, **5**} | {3, 6, 7} | {$a_2, a_3, a_6, a_9, a_{10}, a_{11}$} |
| 7 | (0,1,1) | {1, 3, 4} | {2, 5, 6} | 15 | (0,1,1,**1**) | {1, 3, 4, **5**}$^D$ | | |
| 8 | (1,1,1) | {1, 2, 3, 4} | {5, 6} | 16 | (1,1,1,**1**) | {1, 2, 3, 4, **5**} | {6, 7} | {$a_7, a_9, a_{10}, a_{11}$} |

**Table 9.** Results in the last iterative for vectors 17-32 based on that of 1-16, respectively.

| $i$ | $X_i$ | $S(X_i)$ | $U(X_i)$ | $i$ | $X_i$ | $S(X_i)$ | $C(X_i)$ |
|---|---|---|---|---|---|---|---|
| 1 | (0) | {1} | {2, 3} | 17 | (0,0,0,0,**1**) | {1, **6**}$^D$ | |
| 2 | (1) | {1, 2} | {3, 4, 5} | 18 | (1,0,0,0,**1**) | {1, 2, **6**}$^D$ | |
| 3 | (0,1) | {1, 3} | {2, 4, 6} | 19 | (0,1,0,0,**1**) | {1, 3, **6**} | {$a_1, a_3, a_6, a_9, a_{10}, a_{12}$} |
| 4 | (1,1) | {1, 2, 3} | {4, 5, 6} | 20 | (1,1,0,0,**1**) | {1, 2, 3, **6**} | {$a_4, a_6, a_7, a_8, a_{10}, a_{11}$} |
| 5 | (0,0,1) | {1, 4}$^D$ | | 21 | (0,0,1,0,**1**) | {1, 4, **6**}$^D$ | |
| 6 | (1,0,1) | {1, 2, 4} | {4, 5, 6} | 22 | (1,0,1,0,**1**) | {1, 2, 4, **6**}$^D$ | |
| 7 | (0,1,1) | {1, 3, 4} | {2, 5, 6} | 23 | (0,1,1,0,**1**) | {1, 3, 4, **6**} | {$a_4, a_5, a_6, a_9, a_{12}$} |
| 8 | (1,1,1) | {1, 2, 3, 4} | {5, 6} | 24 | (1,1,1,0,**1**) | {1, 2, 3, 4, **6**} | {$a_5, a_8, a_{10}, a_{12}$} |

| | | | | |
|---|---|---|---|---|
| 9 | (0,0,0,1) {1, **5**}$^D$ | | 25 | (0,0,0,1,**1**) {1, 5, **6**}$^D$ |
| 10 | (1,0,0,1) {1, 2, **5**} | {3, 4, 6, 7} | 26 | (1,0,0,1,**1**) {1, 2, 5, **6**}$^D$ |
| 11 | (0,1,0,1) {1, 3, **5**}$^D$ | | 27 | (0,1,0,1,**1**) {1, 3, 5, **6**}$^D$ |
| 12 | (1,1,0,1) {1, 2, 3, **5**} | {4, 6, 7} | 28 | (1,1,0,1,**1**) {1, 2, 3, 5, **6**}$^D$ |
| 13 | (0,0,1,1) {1, 4, **5**}$^D$ | | 29 | (0,0,1,1,**1**) {1, 4, 5, **6**}$^D$ |
| 14 | (1,0,1,1) {1, 2, 4, **5**} | {3, 6, 7} | 30 | (1,0,1,1,**1**) {1, 2, 4, 5, **6**}$^D$ |
| 15 | (0,1,1,1) {1, 3, 4, **5**}$^D$ | | 31 | (0,1,1,1,**1**) {1, 3, 4, 5, **6**}$^D$ |
| 16 | (1,1,1,1) {1, 2, 3, 4, **5**} {6, 7} | | 32 | (1,1,1,1,**1**) {1, 2, 3, 4, 5, **6**} {$a_{11}$, $a_{12}$} |

### 5.3 Edge Nodes for Feasible Vectors

$O(|S(X)|)$ and $O(|T(X)|)$ are used to verify the connectivity of $G(S(X))$ and $G(T(X))$ to determine whether $C(X)$ is an MC for any new vector $X$. Let $X$ be an $(n–2)$-tuple feasible binary vector. A new concept called the edge node subset $U(X)$ is proposed, such that node $t$ is an edge node if and only if $e_{s,t} \in E$, and nodes $s$, $t$ are in $S(X)$ and $T(X)$, respectively, that is, an edge node is a node at the edges of MC $C(X)$. For example, in Figure 1, $U(X_{23}) = \{2, 5, 7\}$ and $U(X_{24}) = \{5, 7\}$ if $X_{23} = (0, 1, 1, 0, 1)$ and $X_{24} = (1, 1, 1, 0, 1)$, respectively. Note that $C(X_{23}) = \{a_4, a_5, a_6, a_9, a_{12}\}$ and $C(X_{24}) = \{a_5, a_8, a_{10}, a_{12}\}$.

Let vector $X$ be feasible, then $t \in U(X)$, and $\tau \in [T(X)–U(X)]$. Note that, $X(t) = X(\tau) = 0$. Consider the following two cases to determine whether $X^*$ is feasible, based on its parent $X$:

**CASE 1.** $X^*(v) = \begin{cases} 1 & v = \tau \\ X(v) & \text{otherwise} \end{cases}$

No path exists between any node in $S(X)$ and any node in $[T(X)–U(X)]$. We find that $X^*$ is infeasible. For example, in Figure 1, $X_2 = (1, 0, 0, 0, 0)$ is feasible, and $6 \in [T(X_1)–U(X_1)]$, we have $(1, 0, 0, 0, 1)$ is infeasible. After knowing the edge nodes, the above rule takes $O(1)$ to verify whether both $S(X)$ and $T(X)$ are connected subgraphs and is more efficient than $O(n)$ in Algorithm 3, which is based on PLSA.

**CASE 2.** $X^*(v) = \begin{cases} 1 & v = t \\ X(v) & \text{otherwise} \end{cases}$.

If $G(U(X)–\{t\})$ is connected, both $S(X^*)$ and $T(X^*)$ are connected, that is, $C(X^*)$ is an MC because both $S(X)$ and $T(X)$ are already connected.

By contrast, $C(X^*)$ is not an MC if $G(U(X)–\{t\})$ is disconnected. It takes $O(|U(X)–\{t\}|)$ using

PLSA to verify whether $G(U(X)-\{t\})$ is connected and $O(|U(X)-\{t\}|) << O(|T(X)|)$. Hence, the above method is also more efficient when using PLSA to directly verify the connectivity of $T(X)$.

Based on the above two cases, the proposed edge nodes enhance the efficiency of the PLSA.

**5.4 Isolated Nodes**

Edge nodes can be used to determine whether a vector is feasible or infeasible. To have a more powerful tool to fathom a vector and its offspring rather than just itself, a new concept called isolated nodes is proposed. Node $v$ is called an isolated node in $G(X)$ if $v \in S(X)$ but cannot reach node 1 in $G(S(X))$ or $v \in T(X)$, but cannot reach node $n$ in $G(T(X))$. Note that at least one isolated node exists in $X$ if and only if $G(S(X))$ or $G(T(X))$ are disconnected.

For example, in Figure 5, these dark nodes are in $S(X)$, white nodes are in $T(X)$, and dashed nodes are isolated nodes. Nodes 4, 3, and 2 are the outside, inside, and future isolated nodes in Figure 5(a), 5(b), and 5(c), respectively.

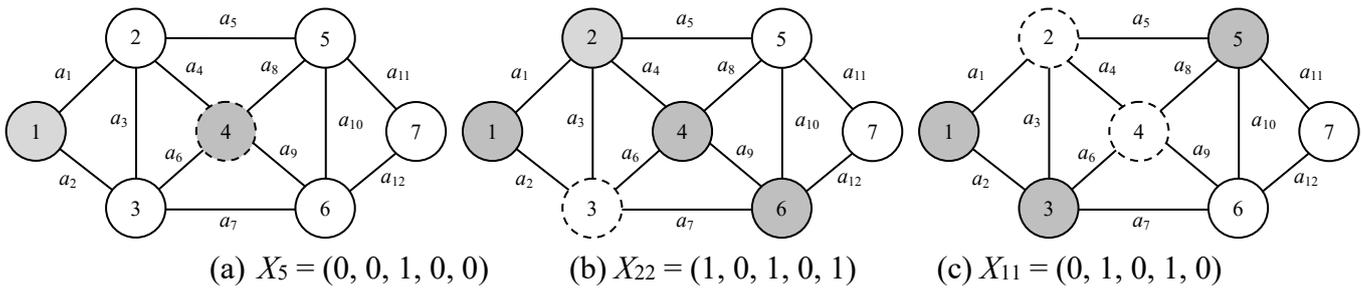

(a) $X_5 = (0, 0, 1, 0, 0)$  (b) $X_{22} = (1, 0, 1, 0, 1)$  (c) $X_{11} = (0, 1, 0, 1, 0)$

**Figure 5.** Example network.

All nodes were numbered in the order of layers obtained from PLSA, as discussed in Section 5.1. It is impossible to reach any node isolated $v$ from 1 or node $n$ in $X$, as shown in Figure 5, that is, $X$ is infeasible if $X$ has an isolated node in $G(X)$. Node $v$ is inside, outside, or future.

In addition, it cannot add any nodes to $S(X)$ such that $v$ is not isolated. Hence, $X$ and all its offspring are infeasible. Moreover, if $X$ is externally isolated, the parent of $X$, that is, $X_1 = (0, 0, 0, 0, 0)$, cannot be included to have any new offspring, such as $(0, 0, 0, 1, 0)$ and $(0, 0, 0, 0, 1)$. Note that $C(X_1)$ and $X_3$ are maintained because $C(X_1)$ is a real MC, and $X_3 = (0, 1, 0, 0, 0)$ is already generated before $X_5 = (0, 0, 1, 0, 0)$.

**5.5 Fathom Vectors based on Isolated Nodes**

Let vector $X$ be feasible and $X_i$ be one son of $X$ such that $X_i(u) = \begin{cases} 1 & \text{if } u = v_i \\ X(u) & \text{otherwise} \end{cases}$ and $X(v_i) = 0$, with $\lambda(v) < \lambda(v_1)$. From the observations in Section 5.3 and 5.4, we have the following rules:

1. Vector $X_1$ is infeasible if node $v_1 \notin U(X)$.

2. Vector $X_2$ and its offspring are infeasible if $X_2(v) = 1$ for all $v \in U(X)$ and $X(u) = 0$ with $\lambda(u) < \lambda(v_2)$, that is, $u$ is an isolated node inside.

3. Vector $X_2$ and its offspring are infeasible if $X_2(v) = 0$ for all $v \in L(\lambda(v))$ and $\lambda(u) < \lambda(v_2)$, that is, $u$ is an outside isolated node. Moreover, the parent of $X$ can be removed from the recursive list.

4. Vector $X_2$ and its offspring are infeasible if $X_2(v_2) = 1$, $X(u) = 0$, $u, v_2 \in U(X)$, $u < v_2$, and $U(X)$ is disconnected, that is $u$ is a future isolated node.

5. Vector $X_2$ is feasible if $v_2 \in U(X)$ and any of the following two conditions is satisfied:

   1) $v_2$ is independent node.

   2) $v_2$ is dependent node and $X(u) = 1$ for all nodes $u$ depended on $v_2$.

Note that these rules not only fathom vector $X$, but also any $X^*$ with $X^*(i) = X(i)$ for $i = 1, 2, \ldots, u$. Hence, these rules can save the runtime in verifying the connectivity of $X$ and all $X^*$, and also the runtime to generate $X^*$ in the BAT. Hence, these rules are implemented in the proposed algorithm to fathom all the related vectors.

## 6 PROPOSED ALGORITHM

The pseudocode, example, and the performance test of the proposed algorithm are provided in this section.

### 6.1 Pseudocode

The pseudocode of the proposed node-based recursive BAT is displayed as follows.

**Algorithm 5: Proposed Node-based Recursive BAT**

**Input:** $G(V, E)$, source node 1 and sink node $n$.

**Output:** All MCs.

**STEP 0.** Find $V(v)$ and $\lambda(v)$ for all $v \in V$ using PLSA; let $i = 2$, $j = 1$, $n^* = (n-2)$, $N = N^* = 2$, $X_1 = (0)$, $X_2 = (1)$, $S(X_1) = \{1\}$, $S(X_2) = \{1, 2\}$, $U(X_1) = V(1)$, $U(X_2) = V(2) - \{1\}$, and $\Omega = \{C(X_1), C(X_2)\}$.

**STEP 1.** Let $X(v) = \begin{cases} 1 & \text{if } v = i \\ X_j(v) = 0 & \text{otherwise} \end{cases}$ and $S(X) = S(X_j) \cup \{(i+1)\}$ based on Eq. (4) and Eq. (5).

**STEP 2.** If there is an isolated node, say $u$, in $G(X)$, $X$ and its offspring are all infeasible, and go to STEP 3; otherwise, go to STEP 4.

**STEP 3.** If $u$ is an outside isolated node, remove $X_j$, and go to STEP 4.

**STEP 4.** If node $(i+1) = 3 \in U(X_j)$, let $U(X) = U(X_j) \cup V((i+1)) - S(X)$, $\Omega = \Omega \cup \{C(X)\}$, $N = N + 1 = 3$, and $X_N = X$.

**STEP 5.** If $j < N^*$, let $j = j + 1$ and go to STEP 1.

**STEP 6.** If $i < n^*$, let $i = i + 1$, $j = 1$, $N^* = N$, and go to STEP 1. Otherwise, halt and $\Omega$ is a complete MC set.

In the pseudocode, STEP 0 implements the PLSA to calculate $V(v)$ and $\lambda(v)$ for all $v \in V$. STEP 1 follows the concept in the iterative BAT proposed in [25] to find all vectors. However, $X$ is changed from an arc-based $m$-tuple vector to a node-based $n^*$-tuple vector. STEPs 2-4 verify the connectivity of $S(X)$ based on the layer number and properties proposed in Section 5.1.

The major time complexity for the proposed algorithm is to find all feasible vectors, which is $O(2^k)$, based on the iterative BAT in finding all $k$-tuple binary-state vectors. STEPs 2-4 take $O(n-2)$ to verify whether the obtained vector is feasible. Hence, the time complexity of the above algorithm is $O((n-2)2^{(n-2)}) = O(n2^n)$ when finding all MCs in binary-state networks.

## 6.2 Step-by-Step Example

It is NP-hard and #P-hard to enumerate all the MCs and calculate the exact binary-state network reliability in terms of MCs in binary-state networks [13]. Hence, the computational cost increases exponentially with network size. Figure 1 is one of the popular benchmarks of binary-state networks [10-19] and its size is very suitable for explaining the proposed algorithm step by step as follows:

Table 10. The BAT and the IET terms

| $v$ | $V(v)$ | $\lambda(v)$ |
|---|---|---|
| 1 | {2, 3} | 1 |
| 2 | {3, 4, 5} | 2 |
| 3 | {2, 4, 6} | 2 |
| 4 | {5, 6} | 3 |
| 5 | {4, 6, 7} | 3 |
| 6 | {4, 5, 7} | 3 |
| 7 | {5, 6} | 4 |

**STEP 0.** Find $V(v)$ and $\lambda(v)$ for all $v \in V$ as shown in Tables 5 and 6; let $i = 2, j = 1, n^* = (n-2) = 5$, $X_1 = (0), S(X_1) = \{1\}, U(X_1) = V(1), X_2 = (1), S(X_2) = \{1, 2\}, U(X_2) = V(2) - \{1\} = \{3, 4, 5\}$, $N = N^* = 2$, and $\Omega = \{ C(X_1) = \{a_1, a_2\}, C(X_2) = \{a_2, a_3, a_4, a_5\} \}$.

**STEP 1.** Let $X(v) = \begin{cases} 1 & \text{if } v = i \\ X_j(v) = 0 & \text{otherwise} \end{cases}$, i.e., $X = (0, 1)$, and $S(X) = S(X_j) \cup \{(i+1)\} = \{1, 3\}$.

**STEP 2.** Because no isolated nodes in $G(X)$, go to STEP 4.

**STEP 4.** Because node $(i+1) = 3 \in U(X_j)$, $X$ is feasible and let $U(X) = U(X_j) \cup V((i+1)) - S(X) = \{2, 3\} \cup \{1, 2, 4, 6\} - \{1, 3\} = \{2, 4, 6\}$, $\Omega = \Omega \cup \{C(X)\} = \{ \{a_1, a_2\}, \{a_2, a_3, a_4, a_5\}, \{a_1, a_3, a_6, a_7\} \}$, $N = N + 1 = 3$, and $X_N = X$.

**STEP 5.** Because $j = 1 < N^* = 2$, let $j = j + 1 = 2$, and go to STEP 1.

**STEP 1.** Let $X(v) = \begin{cases} 1 & \text{if } v = 2 \\ X_2(v) = 0 & \text{otherwise} \end{cases}$, i.e., $X = (1, 1)$, and $S(X) = S(X_2) \cup \{(2+1)\} = \{1, 2, 3\}$.

**STEP 2.** Because no isolated nodes in $G(X)$, go to STEP 4.

**STEP 4.** Because node $(i+1) = 3 \in U(X_j)$, let $U(X) = U(X_2) \cup V(3) - S(X) = \{4, 5, 6\}$, $\Omega = \Omega \cup \{C(X)\} = \{ \{a_1, a_2\}, \{a_2, a_3, a_4, a_5\}, \{a_1, a_3, a_6, a_7\}, \{a_4, a_5, a_6, a_7\} \}$, $N = N + 1 = 4$, and $X_N = X_4 = X$.

**STEP 5.** Because $j = N^* = 2$, go to STEP 6.

**STEP 6.** Because $i = 2 < n^* = 5$, let $i = i + 1 = 3, j = 1, N^* = N = 4$, and go to STEP 1.

**STEP 1.** Let $X(v) = \begin{cases} 1 & \text{if } v = i \\ X_j(v) = 0 & \text{otherwise} \end{cases}$, i.e., $X = (0, 0, 1)$, and $S(X) = S(X_j) \cup \{(i+1)\} = \{1, 4\}$.

**STEP 2.** Because node 4 is isolated, $X$ and its offspring are all infeasible and go to STEP 3.

**STEP 3.** Because node 4 is isolated outside, remove $X_1$ and go to STEP 5.

**STEP 5.** Because $j = 1 < N^* = 4$, let $j = j + 1 = 2$ and go to STEP 1.

$$\vdots$$

Assume that $i = 5$, $j = 12$, $N^* = 10$, and $N = 4$.

**STEP 1.** Let $X(v) = \begin{cases} 1 & \text{if } v = i \\ X_j(v) = 0 & \text{otherwise} \end{cases}$, i.e., $X = (1, 1, 1, 1, 1)$, and $S(X) = S(X_j) \cup \{(i+1)\} = \{1, 2, 3, 4, 5, 6\}$.

**STEP 2.** Because no isolated nodes in $G(X)$, go to STEP 4.

**STEP 4.** Because node $(i+1) = 6 \in U(X_j)$ is connected, let $U(X) = U(X_2) \cup V(3) - S(X) = \{7\}$, $\Omega = \Omega \cup \{C(X)\}$, $N = N + 1 = 5$, $X_N = X$, and go to STEP 5.

**STEP 5.** Because $j = N^* = 12$, go to STEP 6.

**STEP 6.** Because $i = n^* = 5$, halt and all MCs are found in $\Omega = \{\{a_1, a_2\}, \{a_2, a_3, a_4, a_5\}, \{a_1, a_3, a_6, a_7\}, \{a_4, a_5, a_6, a_7\}, \{a_2, a_3, a_5, a_6, a_8, a_9\}, \{a_1, a_3, a_4, a_7, a_8, a_9\}, \{a_5, a_7, a_8, a_9\}, \{a_2, a_3, a_4, a_8, a_{10}, a_{11}\}, \{a_4, a_6, a_7, a_8, a_{10}, a_{11}\}, \{a_2, a_3, a_6, a_9, a_{10}, a_{11}\}, \{a_7, a_9, a_{10}, a_{11}\}, \{a_1, a_3, a_6, a_9, a_{10}, a_{12}\}, \{a_4, a_6, a_7, a_8, a_{10}, a_{11}\}, \{a_4, a_5, a_6, a_9, a_{12}\}, \{a_5, a_8, a_{10}, a_{12}\}, \{a_{11}, a_{12}\}\}$.

The complete results obtained from the proposed algorithm are presented in Tables 5-9.

**6.3 Computation Experiments**

To validate the effectiveness and efficiency of the proposed algorithm, it is necessary to examine it on benchmark networks. Hence, the most popular 20 binary-state benchmark networks, as illustrated in Figures 6(1)-7(20) [13-40], were conducted in the experiment to test the performance of the proposed recursive node-based BAT.

Note that the order of node numbers is renumbered based on Section 5.1, e.g., Figures 6(13) and 6(18) are redrawn to Figures 6(21) and 6(22), respectively.

The performance of the proposed algorithm is compared with that of the best-known node-based algorithm, which is rewritten in BAT, as described in Section 4. For a fair comparison, both algorithms are coded in DEV C++ 5.11 under 64-bit Windows 10 with Intel Core i7-6650U @ 2.20GHz 2.21GHz and 32 GB RAM. Moreover, Pr($a$) = 0.9 for each arc $a$ in all benchmark networks.

The experimental results are listed in Table 11. The best values among the proposed algorithm and node-based MC algorithm are denoted in bold. The notations $T_{NBA}$ and $T_{RBAT}$ are the number and the related runtime required to generate vectors for the best-known node-based algorithm [] and the proposed algorithm, respectively.

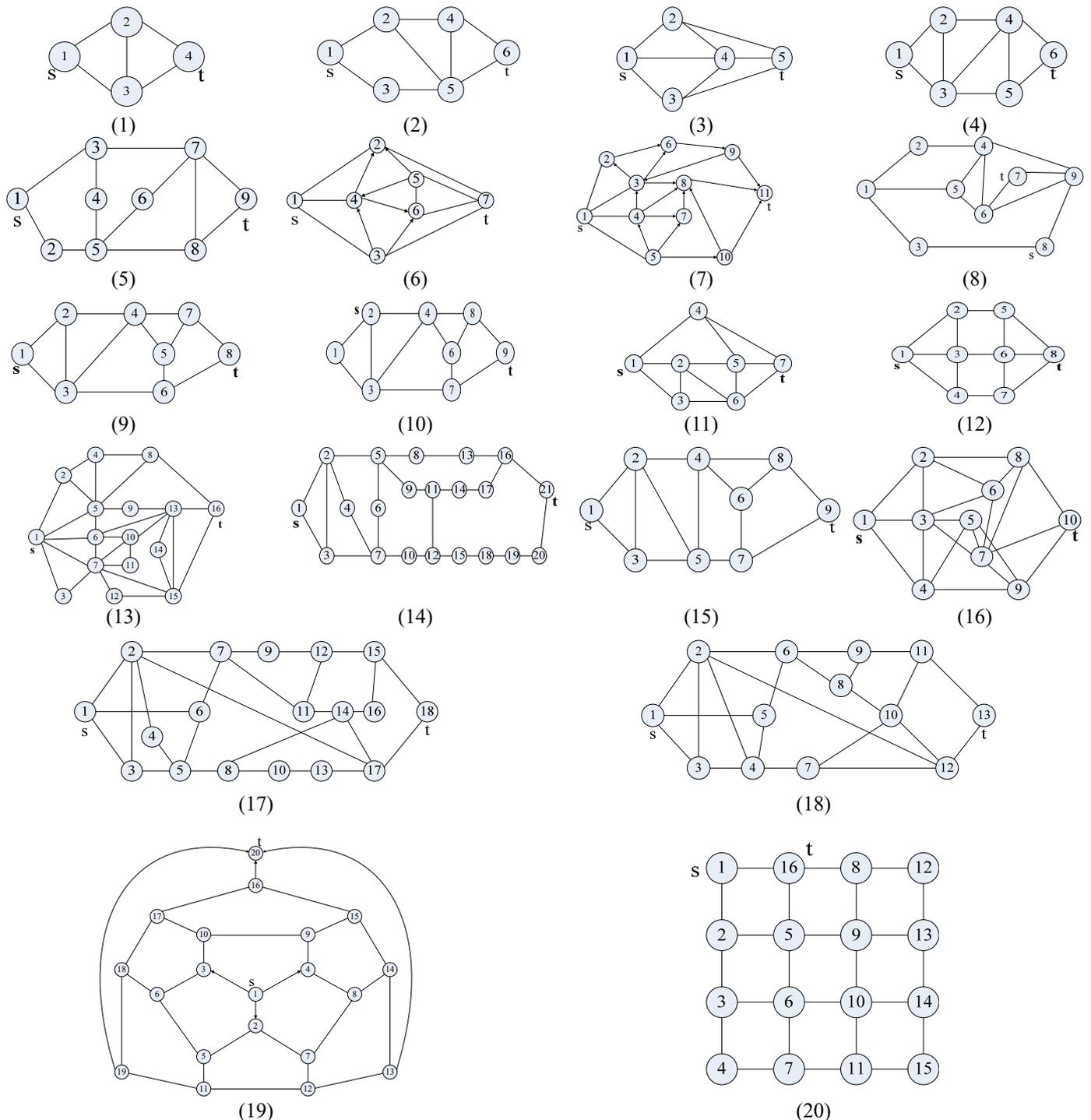

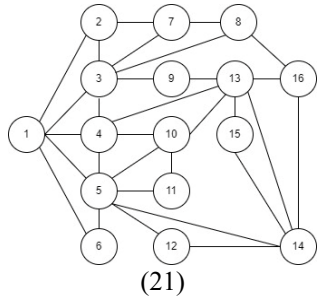
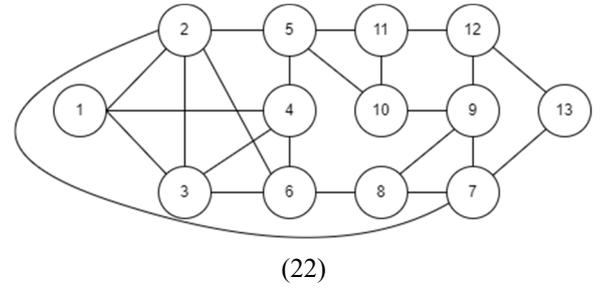

(21) (22)

**Figure 6.** Test networks.

**Table 11.** Comparison of results.

| Fig. | n | m | c | T$_{NBA}$ | T$_{RBAT}$ |
|---|---|---|---|---|---|
| 1 | 4 | 5 | 4 | 0 | 0 |
| 2 | 6 | 8 | 9 | 0 | 0 |
| 3 | 5 | 8 | 8 | 0 | 0 |
| 4 | 6 | 9 | 9 | 0 | 0 |
| 5 | 9 | 12 | 28 | 0 | 0 |
| 6 | 7 | 14 | 25 | 0 | 0 |
| 7 | 11 | 21 | 133 | 0 | 0 |
| 8 | 9 | 13 | 24 | 0 | 0 |
| 9 | 8 | 12 | 19 | 0 | 0 |
| 10 | 9 | 14 | 25 | 0 | 0 |
| 11 | 7 | 12 | 20 | 0 | 0 |
| 12 | 8 | 13 | 29 | 0 | 0 |
| 13 | 16 | 30 | 644 | 0.083 | 0.012 |
| 14 | 21 | 26 | 528 | 0.041 | 0.018 |
| 15 | 9 | 14 | 25 | 0 | 0 |
| 16 | 10 | 21 | 104 | 0 | 0 |
| 17 | 18 | 27 | 1249 | 0.159 | 0.022 |
| 18 | 13 | 22 | 214 | 0.003 | 0.001 |
| 19 | 20 | 30 | 7376 | 0.436 | 0.093 |
| 20 | 16 | 24 | 105 | 0.001 | 0.001 |

The number of obtained MCs increase exponentially with the number of nodes $n$, as shown in Table 12. The above observations satisfy the characteristics of NP-hard and #P-hard problems.

For small problems, such as Figures 6(1)-6(12), which can be solved within 0.001 second, the performance does not seem to differ between the two algorithms. However, the proposed algorithm outperformed the MC algorithm proposed in [19]. The main reason can also be explained by time complexity, as shown below:

**Table 12.** Comparison of the proposed algorithm and the one proposed in [19].

|  | The proposed algorithm | The algorithm proposed in [19] |
|---|---|---|
| Generation | $O(1)$ | $O(n)$ |
| Feasibility Verification | $O(1)$ | $O(n)$ |
| Fathom offspring | Yes | No |

Hence, from both Tables 11 and 12, the proposed node number reorder and isolated nodes reduce the number of infeasible vectors, the edge nodes speed up the procedure to verify the feasibility of vectors, and the recursive algorithm generates vectors efficiently.

The foregoing experimental results are consistent with the comparison of the time complexity between the proposed algorithm and the best-known MC algorithm and further reveal the advantage of the proposed algorithm over the best-known MC algorithm proposed in [].

# 7. CONCLUSIONS

Network reliability is an important index that reflects the status probability of the current networks. Conditional network reliability problems that consider real-life limitations are more practical than those that do not. Before evaluating conditional network performance, MCs and MPs must be present. The recursive BAT improves the efficiency of the traditional BAT, which is the most straightforward implicit enumeration method for having all possible vectors, including MCs. Thus, a new recursive node-based BAT was proposed to search for all MCs for conditional network reliability problems.

The proposed algorithm is the first recursive algorithm and the first BAT to find all MCs. It integrates the recursive BAT to generate each vector in $O(1)$, renumber nodes to fathom infeasible vectors and all their offspring using the isolated nodes to reduce the number of infeasible vectors, and the concept of edge nodes to reduce the runtime in verifying the feasibility of vectors. From a comprehensive comparison with the best-known node-based MC algorithm from both the time complexity and experiments conducted on 20 benchmark binary-state networks, the performance of the proposed recursive node-based BAT was validated.

To establish these benefits, the proposed algorithm will be tested using large benchmarks in the future. In future work, the proposed algorithm will be extended to find MCs, called $d$-MCs, in MFN problems.

# ACKNOWLEDGEMENTS

This research was supported in part by the Ministry of Science and Technology, R.O.C. under

grant MOST 107-2221-E-007-072-MY3 and MOST 110-2221-E-007-107-MY3. This article was once submitted to arXiv as a temporary submission that was just for reference and did not provide the copyright.